\documentclass[aip, amsmath,amssymb,
preprint]{revtex4-1}
\usepackage{graphicx}
\usepackage{dcolumn}
\usepackage{bm}

\usepackage[utf8]{inputenc}
\usepackage[T1]{fontenc}
\usepackage{mathptmx}

\draft 

\begin{document}

\title[Efficient parametric frequency conversions in lithium niobate nanophotonic chips]{Efficient parametric frequency conversions in lithium niobate nanophotonic chips} 

\author{Jia-yang Chen}
\affiliation{Department of Physics and Engineering Physics, Stevens Institute of Technology,  1 Castle Point Terrace, Hoboken, New Jersey 07030, USA}
\affiliation{Center for Quantum Science and Engineering, Stevens Institute of Technology, 1 Castle Point Terrace, Hoboken, New Jersey  07030, USA}
\author{Yong Meng Sua}
\affiliation{Department of Physics and Engineering Physics, Stevens Institute of Technology,  1 Castle Point Terrace, Hoboken, New Jersey 07030, USA}
\affiliation{Center for Quantum Science and Engineering, Stevens Institute of Technology, 1 Castle Point Terrace, Hoboken, New Jersey  07030, USA}
\author{Zhao-hui Ma}
\affiliation{Department of Physics and Engineering Physics, Stevens Institute of Technology,  1 Castle Point Terrace, Hoboken, New Jersey 07030, USA}
\affiliation{Center for Quantum Science and Engineering, Stevens Institute of Technology, 1 Castle Point Terrace, Hoboken, New Jersey  07030, USA}
\author{Chao Tang}
\affiliation{Department of Physics and Engineering Physics, Stevens Institute of Technology,  1 Castle Point Terrace, Hoboken, New Jersey 07030, USA}
\affiliation{Center for Quantum Science and Engineering, Stevens Institute of Technology, 1 Castle Point Terrace, Hoboken, New Jersey  07030, USA}
\author{Zhan Li}
\affiliation{Department of Physics and Engineering Physics, Stevens Institute of Technology,  1 Castle Point Terrace, Hoboken, New Jersey 07030, USA}
\affiliation{Center for Quantum Science and Engineering, Stevens Institute of Technology, 1 Castle Point Terrace, Hoboken, New Jersey  07030, USA}
\author{Yu-ping Huang}
\affiliation{Department of Physics and Engineering Physics, Stevens Institute of Technology,  1 Castle Point Terrace, Hoboken, New Jersey 07030, USA}
\affiliation{Center for Quantum Science and Engineering, Stevens Institute of Technology, 1 Castle Point Terrace, Hoboken, New Jersey  07030, USA}


\date{\today}

\begin{abstract}
Chip-integrated nonlinear photonics holds the key for advanced optical information processing with superior performance and novel functionalities. Here, we present an optimally mode-matched, periodically poled lithium niobate nanowaveguide for efficient parametric frequency conversions on chip. Using a 4-mm nanowaveguide with subwavelength mode confinement, we demonstrate second harmonic generation with efficiency over $2200\%~W^{-1}cm^{-2}$, and broadband difference frequency generation with similar efficiency over a 4.5-THz spectral span. These allow us to generate correlated photon pairs over multiple frequency channels via spontaneous parametric down conversion, all in their fundamental spatial modes, with a coincidence to accidental ratio as high as 600. The high efficiency and dense integrability of the present chip devices may pave a viable route to scalable nonlinear applications in both classical and quantum domains. 
\end{abstract}


\maketitle 

\section{Introduction}
Second order optical nonlinearities ($\chi^{(2)}$) are central to a plethora of classical and quantum optics studies and applications. In particular, optical parametric processes such as second harmonic generation (SHG), quantum frequency conversion, and parametric down conversion are quintessence for many quantum photonics techniques in computing \cite{Lenzinieaat9331}, communications \cite{Sansoni2017,doi:10.1063/1.5038186}, and metrology \cite{PhysRevLett.117.110801}. Over the years, lithium niobate (LN) has been a preferred material of choice for those applications due to its large  $\chi^{(2)}$ tensor ($d_{33}\sim$27 pm/V), ultrafast electro-optic responses, and a low-loss optical window across from 0.35 $\mu$m to 5.2 $\mu$m. Yet typical lithium niobate devices take the forms of bulk crystal or weakly confined waveguides \cite{Parameswaran:02}, which are challenging for dense integration, as needed for hosting complex functionalities. Furthermore, the weak mode confinement limits the achievable interaction and thus requires high laser power to obtain strong nonlinear effects. 

Quite recently, the advent of monolithic LN integrated photonics with high index contrast and low loss points to integrated nonlinear devices on chip \cite{Wang2018}. With strong field enhancement by subwavelength mode confinement, they promise to overcome the aforementioned deficiencies, leading to efficient devices with much enhanced nonlinear interaction strength, smaller footprints, and highly scalable \cite{Wang:18}. However, the phase matching requirement for efficient nonlinear processes is stringent in LN nanostructures, which hinders the development of integrated nonlinear photonic chips. Several approaches have been put forward to achieve phase matching, including by utilizing birefringent and geometric dispersion \cite{PhysRevA.98.023820,zenojyc} or higher-order mode interactions \cite{Chen:18,doi:10.1002/lpor.201800288}. Nonetheless, both approaches are fundamentally limited, where the former cannot access the $d_{33}$ tensor with the highest $\chi^{(2)}$ susceptibility, while latter suffers the poor mode overlapping between the fundamental and higher-order modes. In contrast, by periodically inverting the optical crystal domains, quasi phase matching can be achieved among fundamental spatial modes while taking advantage of the $d_{33}$ tensor element, both crucial for strong nonlinear interactions. In addition, this facilitates the interfacing of LN chips with other devices, such as quantum emitters \cite{Sipahigil847, doi:10.1063/1.5054865} and optical fibers, thereby allowing interconnection and reconciliation of quantum resources on disparate platforms towards practical deployment of scalable quantum network \cite{Wehnereaam9288,2019arXiv190109826K,2018arXiv180504011L}.

In this paper, we design and fabricate periodically poled lithium niobate (PPLN) nanowaveguides with carefully tailored quasi-traverse-electric (TE$_{00}$) mode profile and achieving subwavelength light confinement for efficient parametric frequency conversions. This allow us to harness the highest $\chi^{(2)}$ nonlinear susceptibilities while benefiting from maximum modes overlap of interacting wavelengths and strong light confinement. We observe high conversion efficiency of $2200\%~W^{-1}cm^{-2}$, which is comparable to the record efficiency on nanophotonic PPLN\cite{Wang:18}. Furthermore, we demonstrate difference frequency generation (DFG), stimulated analogues of spontaneous parametric down conversion (SPDC) covering more than 4.5 THz in telecom band, currently limited only by the accessible wavelength range of the our laser. By type-0 quasi-phase matched SPDC, we generate correlated photon pairs in multiple spectral channels, each in a single fundamental spatial mode,  and observe coincidence to accidental ratio (CAR) as high as 600. 

\section{PPLN Waveguide Design and Fabrication}

The efficiency of parametric frequency conversions in PPLN nanowaveguides depends on the nonlinear susceptibility tensor, wave vector mismatch, spatial mode confinement and overlap of the interacting lights \cite{Parameswaran:02}. Taking those factors into account, the normalized conversion efficiency is given as,
\begin{equation}
   \eta_{norm} = \frac{8\pi^2}{\epsilon_0 c \lambda^2_{2\omega}} \,  \frac{d_{eff}^2}{n^{2\omega}_{eff} (n^{\omega}_{eff})^2} \, \frac{\iint\,E_{2\omega}^{*}\,E_{\omega}^2\,dx\,dz}{\sqrt{\iint\,|E_{2\omega}|^2\,dx\,dz} \,\iint\,|E_{\omega}|^2\,dx\,dz}\,sinc^2(\Delta K L/2),   
   \label{eq1}
 \end{equation}
where $c$ is the speed of light in vacuum, $\epsilon_0 $ is the vacuum permittivity, $d_{eff}$ is the effective nonlinear coefficient equal $\frac{2}{\pi}\,d_{33}$ in our case, $\Delta K$ 
is the wave vector mismatch, $L$ is interaction length, $n^{(\omega,2\omega)}_{eff}$ and $E_{(\omega,2\omega)}(x,z)$ are the effective indices and electrical fields of 1550-nm quasi-TE$_{00}$ and 775-nm quasi-TE$_{00}$ modes, respectively. We design the waveguide's cross-sectional geometry to realize strongly confined yet optimally overlapping fundamental spatial modes for all interacting lights, as permitted by poling. Specifically, we perform simulations to determine its ideal width and height for subwavelength mode confinement and over $90\%$ overlap. These are crucial in ensuring high nonlinear efficiency while eliminating  cross-mode interaction \cite{Chen:18,Luo:18}, which may prove necessary for achieving single photon nonlinearities in the future \cite{PhysRevLett.113.173601}. Besides, the use of fundamental spatial modes makes it easy for interfacing the current nanophotonic chips with other devices, such as coupling the single photon from the chips to single mode fibers (SMFs) for long distance quantum state transferring. In contrast, coupling higher order modes into a SMF is rather inefficient. 
Furthermore, we periodically pole the waveguide to realize type-0 quasi-phase matching (QPM), with each interacting light guided in a TE mode to fully utilize the largest nonlinearity tensor $d_{33}$ available on the X-cut LN thin film. For SHG, the QPM condition reads

\begin{equation}
 \Delta K=k_{s} - 2k_{p} - \frac{2\pi}{\Lambda_{shg}} = 0, 
\end{equation}
with $2\omega_{p} =\omega_{s}$. 
For SPDC, it reads
\begin{equation}
\Delta K =k_{p} - k_{s} - k_{i} -\frac{2\pi}{\Lambda_{spdc}} = 0,
\end{equation}
with $ \omega_{p} = \omega_{s} +\omega_{i}$.
Here, $k_j=n_{eff}(\omega_j,T) \omega_j/c$ is the wavenumber for the $j$-th lightwave, with $j=p,s,i$ corresponding to the pump, signal, and idler, respectively. $n$ and $\omega_j$ are effective refractive index and angular frequency. $T$ and $\Lambda$ are the temperature and poling period of the PPLN. 

By using the temperature-dependent generalized Sellmeier equation and the Lumerical MODE solution to determine $n_{eff}$ for interacting wavelengths in a mode matched nanowaveguide \cite{Chen:18}, we find that $k_s +k_i \approx 2 k_p$, over a wide bandwidth in telecom wavelengths (1500- 1600 nm). This implies $k_{p,SHG} = k_{p,SPDC}$ given that $\Lambda_{SHG}=\Lambda_{SPDC}$, allowing us to determine the optimum pump wavelength for SPDC by the means of identifying optimum second harmonic wavelength. Unlike the large period ( $\Lambda > 10 \mu m$) in traditional PPLN waveguides with weakly confined mode, the required poling period for tightly confined modes in LN nanowavguides will be very small ($ \Lambda\sim 4 \mu m$) due to much larger wavevector mismatch in $\Delta k$ \cite{Parameswaran:02}, posing a considerable challenge on the poling process. For this reason, the limitation on poling period is taken into consideration for the optimization of waveguide cross sectional geometries.

\begin{figure}
\includegraphics[width=5in]{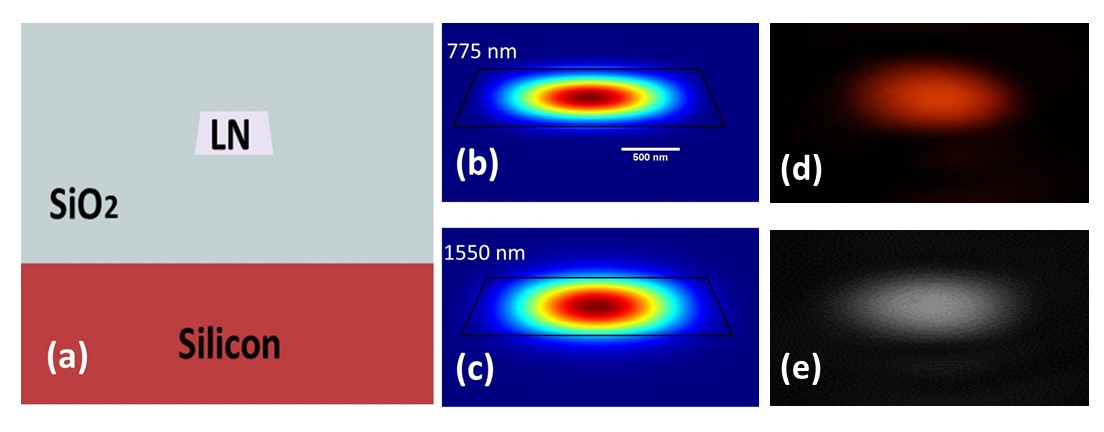}
\caption{\label{fig:figure1}(a) The cross-section of a typical PPLN nanowaveguide. (b)-(c) and (d)-(e) are the simulated and measured mode profiles for 775 nm and 1550 nm quasi-TE$_{00}$ mode, respectively.}
\label{figure1}
\end{figure}

Figure \ref{figure1}(a) shows the optimized cross section of a LN nanowaveguide, with 500 nm in height, 1850 nm in the top width, and a $67^\circ$ sidewall angle. The inclusion of the sidewall angle in our simulation is important for accurately simulating the optical modes to calculate the phase matching curve of the PPLN nanowaveguide. The required poling period for this geometry is about 4 $\mu$m. Figure \ref{figure1}(b) and (c) show the simulated profiles for the quasi-TE$_{00}$ modes at 775 nm and 1550-nm, respectively. The measured profile images are shown in Fig.~\ref{figure1}(d) and (e), captured by using a high NA (0.45) aspheric lens with CMOS and IR cameras, respectively. As seen, these two modes are nearly identical, giving the excellent overlap of the interacting waves, as it is crucial to achieve efficient frequency conversions. Furthermore, the fundamental spatial mode over broad wavelength as seen in Fig.~\ref{figure1}(d) and (e) can be very useful for coupling emission from quantum emitter \cite{doi:10.1063/1.4978204} into PPLN for subsequent frequency conversion\cite{PhysRevLett.109.147404}. The use of such optimized spatial modes may prove crucial in extending the existing quantum technology to the challenging mid-IR regime \cite{Sua2017,Kalashnikov2016}. Theoretically, the normalized SHG efficiency, $\eta_{norm}$, calculated using the actual waveguide parameters is more than $5,000 \%W^{-1}cm^{-2}$. By further light confinement, even higher $\eta_{norm}$ is possible \cite{ChenFIO18}, albeit requiring very short poling period ($\sim$ 2.4 $\mu$m). Nonetheless, recent progress on Z-cut LNOI could be an promising alternative approach for achieving uniform ultra-short poling period over a centimeter length \cite{zcut}.

\begin{figure}
\includegraphics[width=5in]{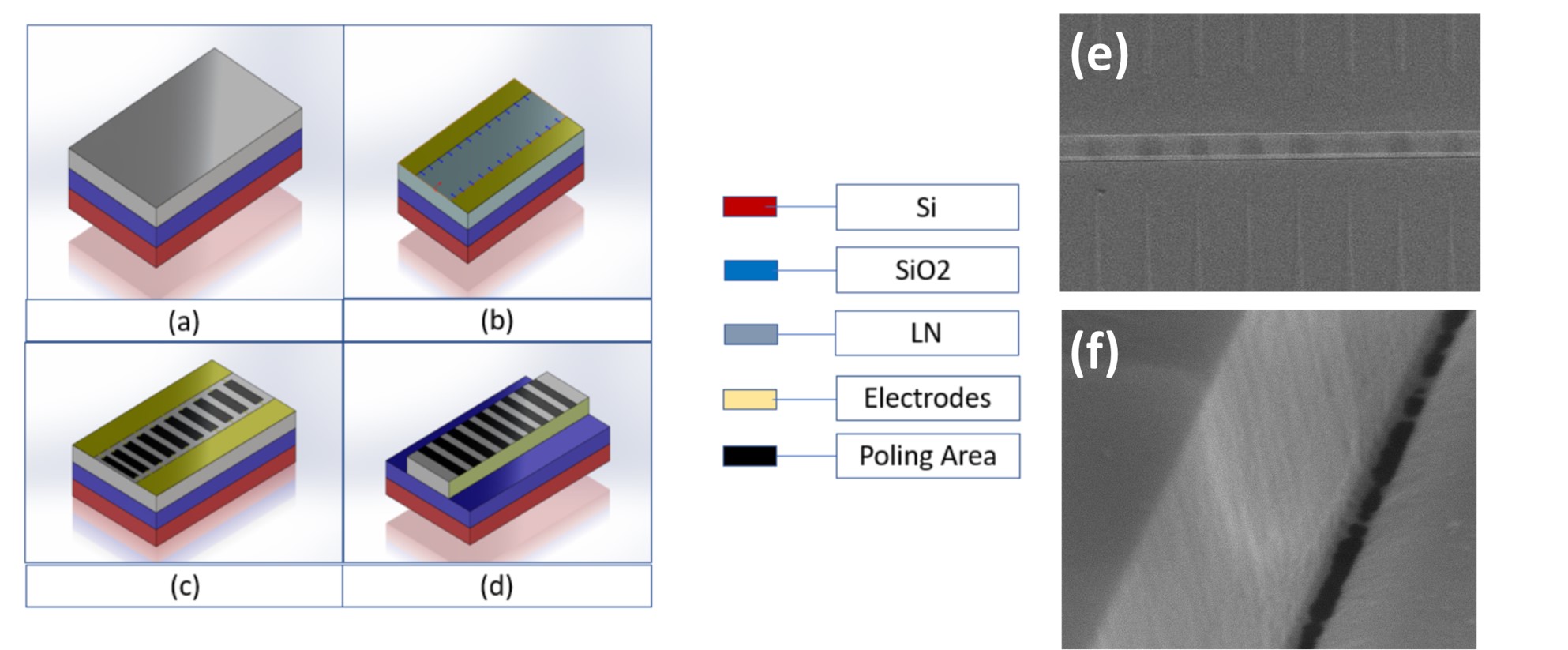}
\caption{\label{fig:figure2}(a) X-cut LNOI wafer. (b) Cr-AU Comb electrodes on LN wafer (c) Poling areas after several 20-ms high voltage pulses. (d) PPLN nanowaveguide after the complete fabrication process. SEM images showing the (e) top view and (f) sidewall of PPLN waveguide before PECVD cladding.}
\label{figure2}
\end{figure}

The PPLN waveguides are fabricated on an X-cut LNOI wafer (NANOLN Inc.), which is a 500 nm thick LN thin film bonded on a 2-$\mu m$ thermally grown silicon dioxide layer above a silicon substrate, shown in  Fig.~\ref{figure2}(a). We use bi-layer electron-beam resist (495 PMMA A4 and 950 PMMA A4) and define the comb-shaped electrodes by using electron-beam lithography (EBL, Elionix ELS-G100, 100 keV). Then 15-nm Cr and 60-nm Au layers are deposited via electron-beam evaporation (AJA e-beam evaporator). The desirable poling electrode pattern is then created by a metal lift-off process. We apply several 20-ms high voltage electrical pulses on the poling pads, shown in Fig.~\ref{figure2}(c), to form the domain inversion region shown in the dark color. Then a second EBL is carried out to define the LN waveguide in the poled region. Using a similar process described in our previous work \cite{Chen:18}, an optimized ion milling process is used to etch the waveguide structure with smooth sidewalls and the optimum sidewall angle. As shown in Fig.~\ref{figure2}(e), we obtain uniform periodic domain inversion, which is crucial for high conversion efficiency. RCA (define) bath for the removal of the redeposition is applied delicately to minimize the sidewall roughness due to the uneven removal rates for poled and unpoled regions. With this, the sidewall roughness of the PPLN waveguide shown in Fig.~\ref{figure2}(f) is nearly unaffected by the cleaning process, giving low propagation loss similar to the normally etched waveguide. To extract the propagation loss, we fabricate a microring resonator with similar waveguide geometry (but with 2 $\mu m$ top width), where the intrinsic quality factor is measured to be two million, indicating the propagation loss $~0.15 dB/cm$ (see the Supplementary Material).

With a smaller poling period, the acute field spreading effect between adjacent teeth become more significant, causing inconsistent domain wall profile. Therefore, in order to maintain the poling quality and acceptable yield for a small poling period ($\sim 4 \mu m$), we use triangle-shape teeth on one side of the electrode while the other side remains in a rectangular shape for grounding. This helps form singly concentrated electrical fields for poling along the teeth direction toward the rectangular electrodes \cite{Reich:98}. This minimizes the lateral spreading of the electric field and improves the poling quality considerably.

\section{Nonlinear process in nanophotonic PPLN waveguides}
\subsection{Second harmonic and difference frequency generations}
\begin{figure}
\includegraphics[width=6.2in]{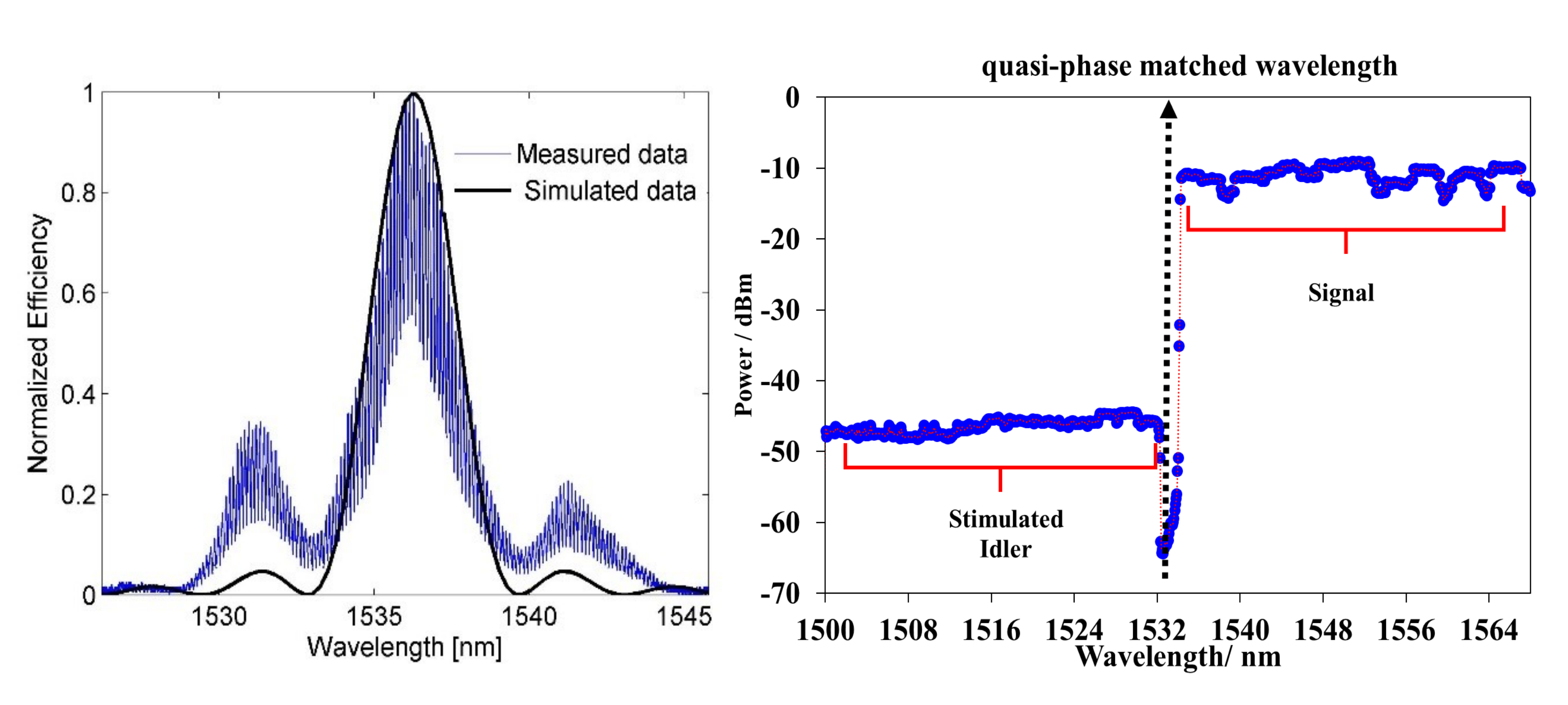}
\caption{\label{fig:figure3}(a) Phase match curve of PPLN waveguide. (b) DFG spectrum when pump at the phase-matched wavelength around 766 nm at 0.1 mW peak power in the waveguide.}
\label{figure3}
\end{figure}

The experimental setup for the SHG efficiency measurement is similar as described in our previous work \cite{ChenFIO18}.
A continuous-wave (CW) tunable laser (Santec 550, 1500-1630 nm) is used as pump laser along with a fiber polarization controller to excite the fundamental quasi-TE mode in the PPLN waveguide.Two tapered fibers (2 $\mu$m spot diameter, OZ optics) serve as the input and output with losses of 4.3 dB per facet at 1550 nm and 5.4 dB per facet at 775 nm, respectively. Fig.~\ref{figure3}(a) shows the measured phase matching curve in the PPLN waveguide at T = 34.5~$^\circ C$. The fitting envelope agrees well with the normalized theoretical phase matching curve indicating decent poling quality and well-satisfied quasi-phase matching. After taking the coupling loss into account, the measured on-chip pump power and SH power are 2.95 mW and 31.56 $\mu$W, respectively. This corresponding to normalized SHG efficiency for this PPLN waveguide with 4 mm length and 4 $\mu$m period of over $2200  \% W^{-1}cm^{-2}$. We believe that lower measured normalized SHG efficiency compared to the theoretical prediction is primarily due to tolerances in device fabrication and poling process. Further improvement on poling process is possible by adding insulation layer such as silicon oxide to prevent current leakage \cite{Nagy:18} hence better poling uniformity. The oscillatory measured phase matching curve in Fig.~\ref{figure3}(a) is caused by the Fabry-Perot (FP) cavity formed between the two polished facets on both end of the waveguide, which can be suppressed by having anti-reflection coating on both facets. Nonetheless, a well defined FP cavity is a noteworthy feature that can be used for preparing high dimensional quantum states \cite{Autebert:17}. Combining highly efficient SHG and light guiding in fundamental spatial mode over extended electromagnetic spectrum may unlock the potential toward chip-scale optical counting or atomic clocks on LN chip, as such, by harnessing its $\chi^{(2)}$ and $\chi^{(3)}$ nonlinearity for octave-spanning supercontinuum generation\cite{2019arXiv190111101Y} and efficient coupling to atomic clocks\cite{PhysRevApplied.8.014027}. 

Then, we characterize the phase matching bandwidth for difference frequency generation (DFG) by using the same PPLN waveguide. In this experiment, the pump and the signal are seeded from optical path 1 and 2 labeled in Fig.~\ref{figure4}, respectively. The pump source is a CW tunable laser (Newport TLB-6712). Then, we fix the pumping wavelength while continuously sweep the signal wavelength (Santec 550) and record the generated idler power spectrum after the long pass filter (LPF) using an optical spectrum analyzer (not shown in Fig.~\ref{figure4}). As shown in Fig.~\ref{figure3}(b), a broadband DFG response ($>$ 4.5 THz) is observed, only limited by the accessible laser wavelength. The broadband DFG measurement promises generating multichannel correlated photon pairs or broadband on-chip parametric amplification with such PPLN nanowaveguides \cite{Sua:18}.

\subsection{Spontaneous parametric down-conversion}

\begin{figure}[htbp]
\includegraphics[width=5in]{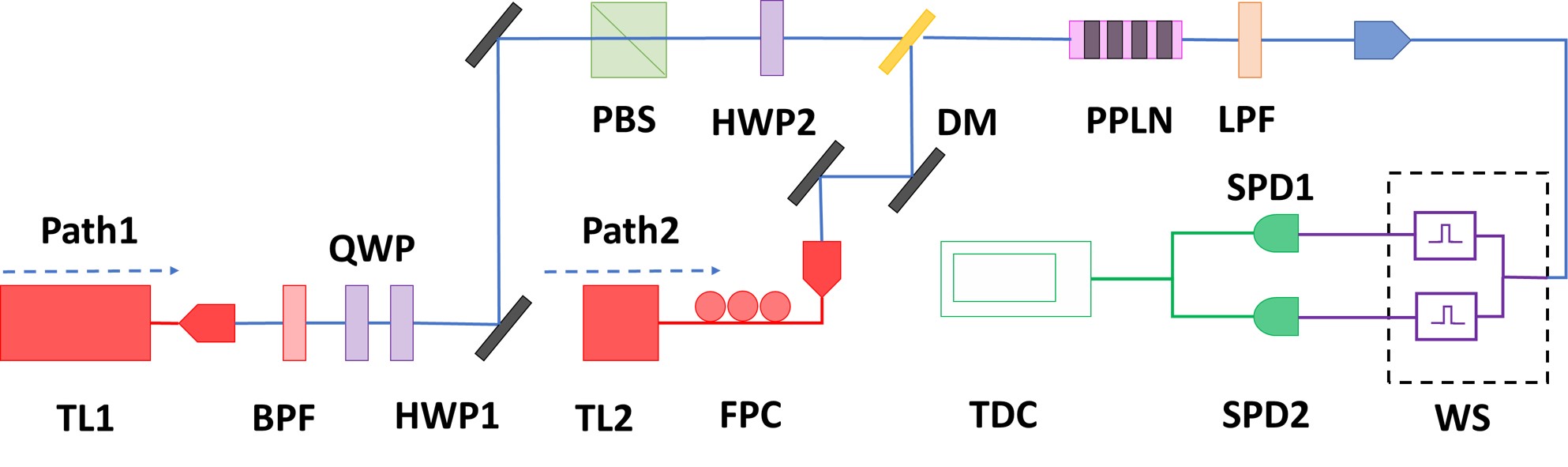}
\centering
 \caption{Schematic of the experimental setup. The pump laser for photon pair generation is obtained from a visible tunable laser (TL1). The pump laser for SHG measurement is a near-infra-red tunable laser (TL2). BPF, band-pass filter, QWP, quarter-wave plate, HWP, half-wave plate, PBS, polarization beam splitter, DM, dichroic mirror, LPF, long-pass filter, SPD, single photon detector, WS, waveshaper, TDC, time-to-digital converter, FPC, fiber polarization controller.}
 \label{figure4}
\end{figure}

Next we examine the performance of this PPLN nanowaveguide for photon pair generation via SPDC. The experimental setup is shown in Fig.~\ref{figure4}. In this experiment, the pump is a CW narrow linewidth ($<$200 KHz) tunable laser around 767 nm (Newport TLB-6712). It is collimated via a fiber collimator and guided through a half-wave plate, a quarter-wave plate and a polarization beam splitter to excite the quasi-TE mode in the waveguide. Then, an additional half-wave plate is used to fine tune the polarization angle prior to coupling into the LN chip by using an AR-coating aspheric lens. The temperature of the PPLN chip is stabilized at 34.5~$^\circ C$ by a heater  with accuracy of 0.1~$^\circ C$. Subsequently, the generated photon pairs will be coupled out from the LN chip by using a second AR-coating aspheric lens. After rejecting the pump with a long-pass filter, the generated photon pairs in the telecom band will be collected into to a SMF-28 optical fiber with 70$\%$ coupling efficiency, achievable thanks to the well-defined fundamental spatial mode of the nanowaveguide. Such single-mode photons are vital for interfacing and manipulation of disparate quantum systems via photon-spin\cite{2018NaPho..12..516A} and photon-quantum dot\cite{2018arXiv181207165P} interfaces. Then the signal and idler photons are separated by two 200 GHz bandpass filters (each over 35 dB extinction ratio) formed by a programmable WaveShaper (FINISAR). The photon pairs are detected by two synchronized, gated InGaAs single photon detectors with 100 MHz repetition rate and 1 ns gate width, where the coincidence counting events are recorded by using a time-to-digital converter \cite{Sua2017}. 

\begin{figure}
\includegraphics[width=6.2in]{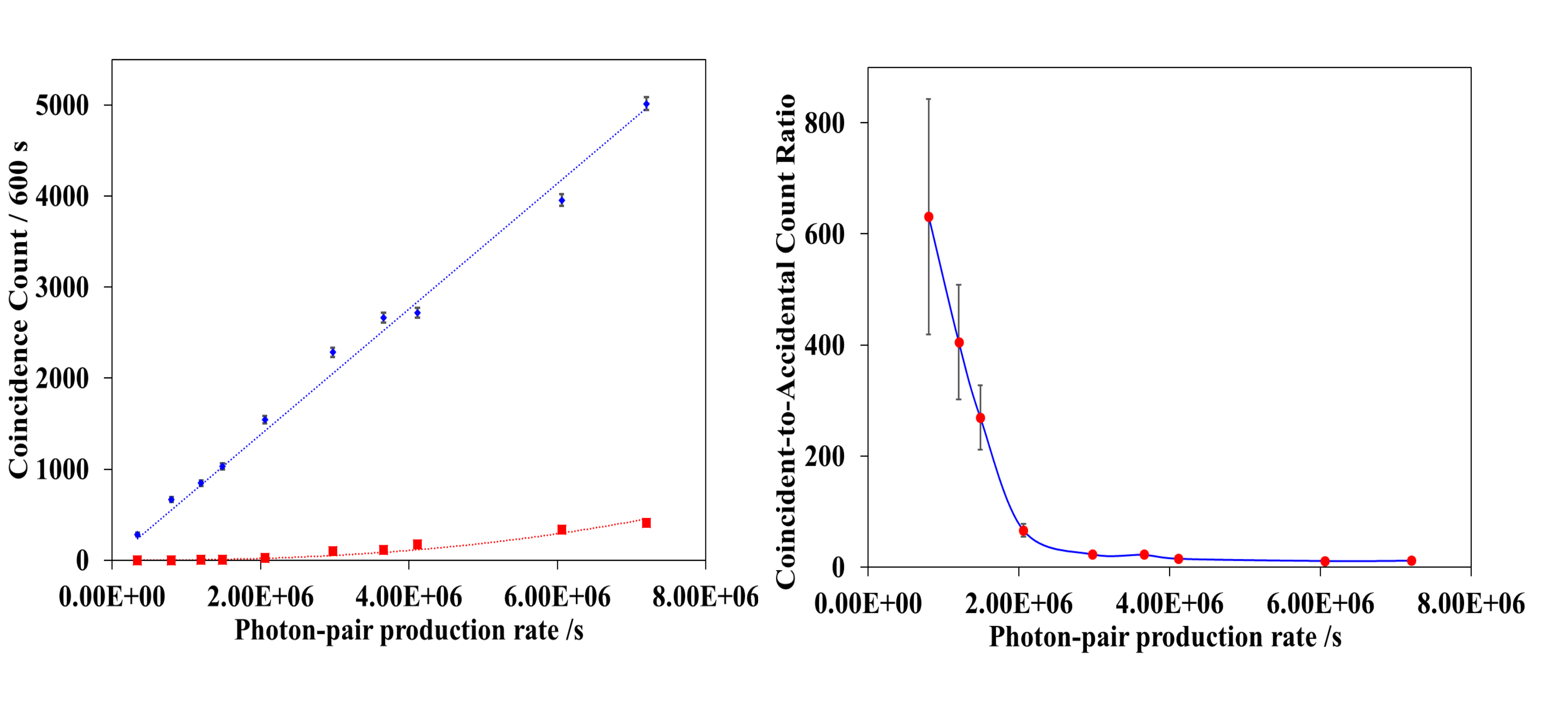}
\centering
\caption{(a)Measured coincidence and accidental counts and (b) coincidence-to-accidental-counts ratio (CAR) as a function of photon production rate.}
 \label{figure5}
\end{figure}

To characterize the correlated photon pairs, we measure the single photon counts, coincidence counts, and accidental-coincidence counts as a function of the pump power to obtain the coincidence-to-accidental-counts ratio (CAR)\cite{Sua2017}. The measured coincidence and accidental-coincidence at a different photon pair production rate is shown in Fig.~\ref{figure5}(a). One can see that the coincidence rate increases linearly with photon production rate while the accidental-coincidence rate increases quadratically. This is because a coincidence event is a result of spontaneous down conversion where one pump photon is annihilated and two daughter (signal and idler) photons are created, as governed by energy conservation in Eq.~\ref{eq1}. Contrarily, the accidental-coincidence counts arise from the coincidence events of uncorrelated photons as a consequence of multiple photon pairs generation, especially with high pump  power. We observe a CAR value of $631 \pm 210$ at photon production rate of 0.8 million pairs per second ($=$ 0.0069 means photon number per gate), corresponding to a brightness of 69 MHz/mW/nm. Note that the CAR value can be improved significantly by reducing the time-frequency detection modes by the means of narrow spectral filtering, time-gating, or the mode selective detection \cite{Shahverdi2017}. 

\begin{figure}[htbp]
\includegraphics[width=5in]{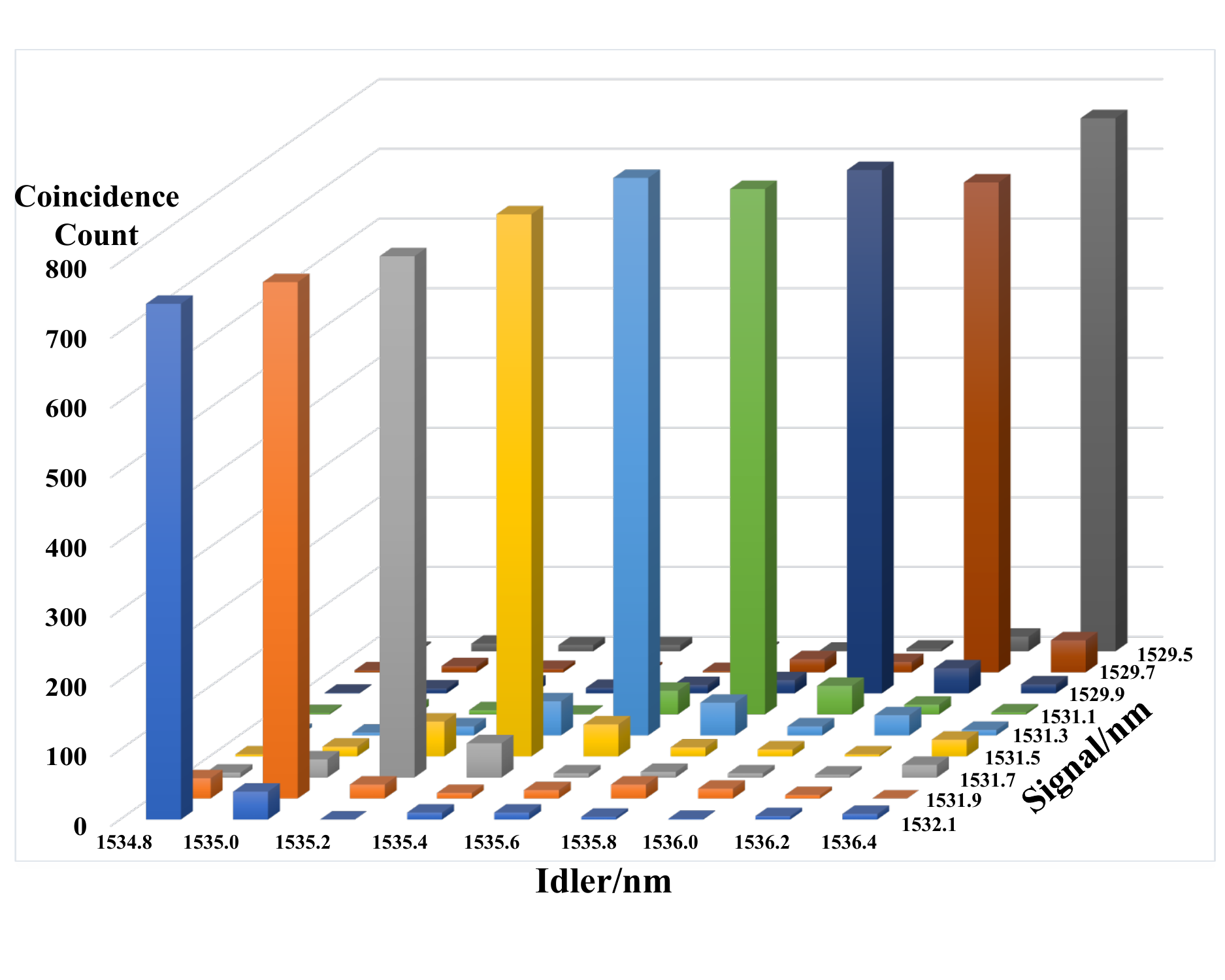}
\centering
 \caption{Coincidence count measured at selected signal/idler wavelength combinations. Significant coincidence counts (corresponding to a peak) are observed only between channels fulfilling energy and momentum conservation. }
 \label{figure6}
\end{figure}

We characterize the spectral correlations between nine combinations of the signal and idler wavelength channels, as allowed by the total bandwidth of our programmable waveshaper as a tunable spectral filter (25 GHz). Here, we observe strong frequency correlations, where consistent photon coincidences were measured only for particular signal-idler combination as shown Fig.~\ref{figure6}, where the diagonal coincidences spectra distribution reflects the energy and  momentum conversations in the SPDC process. Nevertheless, the strong frequency correlation is anticipated to extend over a broad wavelength range as predicted by the observed DFG bandwidth\cite{PhysRevLett.111.193602}, as shown in Fig.~\ref{figure3}(b). Enabling broadband and high purity correlated photon pairs generated on-chip, a potential quantum information application is near-deterministic generation of single photons via spectral multiplexing of heralded single-photon sources\cite{PhysRevLett.119.083601}. Additionally High purity correlated photon source can operate as a stable and accurate central synchronization clock source for  deployment of distant interconnected quantum photonics chips within a quantum network \cite{doi:10.1063/1.1797561,Ho_2009,Ilo-Okeke2018}.

\section{Conclusion}
In summary, we have realized a series of efficient parametric frequency conversions in an optimally mode-matched, periodic-poled lithium niobate nanowaveguide on chip. We first observed second harmonic generation and broadband difference frequency generation over 4.5 THz, whose normalized conversion efficiency is as high as $2200\%~W^{-1}cm^{-2}$, thanks to the tight fundamental mode confinement for all interacting modes and high quality poling. Using the same PPLN nanowaveguide, we implemented a bright, wavelength multiplexing source of correlated photon pairs via spontaneous parametric down conversion, achieving coincidental-to-accidental ratios as high as 600 with only moderate spectral filtering. Highly efficient, optimally fundamental-mode matched, dense integrable, the present frequency conversion devices may find many applications in scalable nonlinear applications in both classical and quantum domains. The next steps will involve incorporating other optical elements on the same lithium-niobate chip, such as electro-optical modulation \cite{Jin:19} wavelength multiplexing, to realize functional chips.

\section*{Appendix A:Loss measurement}
To extract the propagation loss in PPLN waveguide, we fabricate microring resonators with similar dimension on the same type of LNOI wafer and used the same fabrication process. The highest intrinsic Q  $\sim 2\times10{^6}$ is achieved on a 80 $\mu$m radius microring with 2 $\mu$m width. It is indicating the propagation loss is about 0.15 dB/cm.
\begin{figure}
\includegraphics[width=4in]{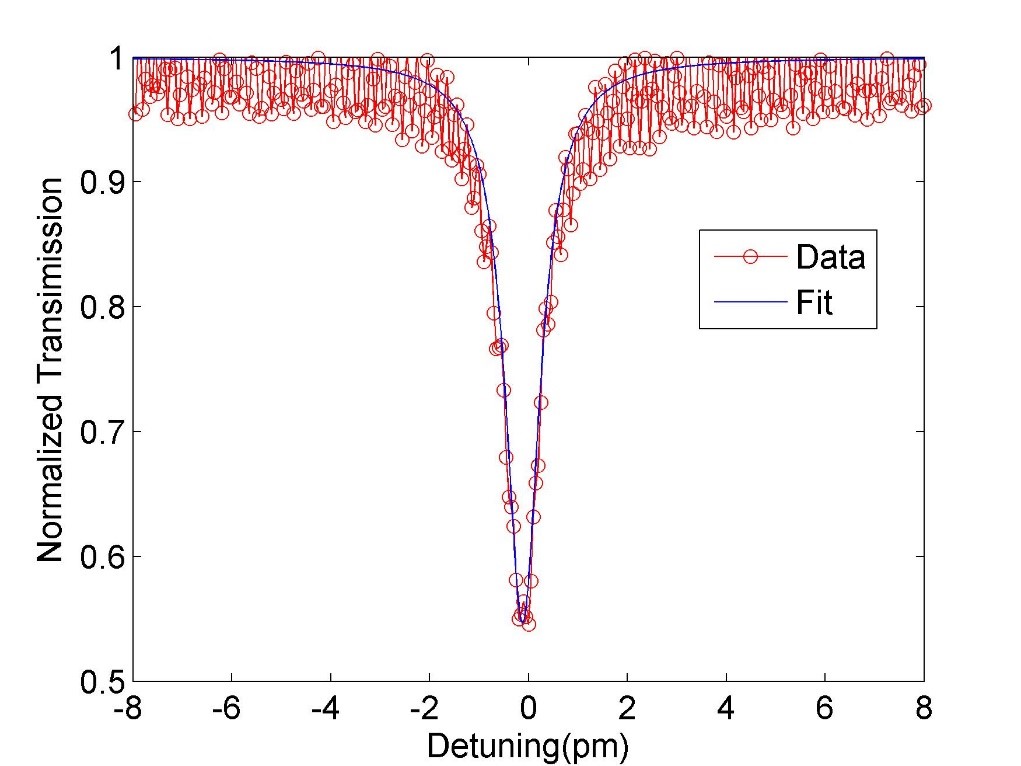}
\centering
 \caption{The spectrum of one mode which yields $2\times10{^6}$ quality factor at around 1600 nm.}
 \label{figure}
\end{figure}

\begin{acknowledgments}
This work was supported in part by the National Science Foundation (Grant No. 1641094 and No. 1842680). This fabrication was performed in part at the Advanced Science Research Center Nanofabrication Facility of the Graduate Center at the City University of New York.
\end{acknowledgments}

\begin{newpage}
\end{newpage}

%

\end{document}